# Bright room temperature single photon source at telecom range in cubic silicon carbide

Junfeng Wang,[1*] Yu Zhou,[1*] Ziyu Wang,[1] Abdullah Rasmita,[1] Jianqun Yang,[2] Xingji Li,[2] Hans Jürgen von Bardeleben,[3] and Weibo Gao[1,4]

[1]*Division of Physics and Applied Physics, School of Physical and Mathematical Sciences, Nanyang Technological University, Singapore 637371, Singapore*

[2]*School of Materials Science and Engineering, Harbin institute of Technology, Harbin 150001, P.R.China*

[3]*INSP, Université Pierre et Marie Curie, UMR 7588 au CNRS 4 place Jussieu, 75005 Paris, France*

[4]*The Photonics Institute and Centre for Disruptive Photonic Technologies, Nanyang Technological University, 637371 Singapore, Singapore*

## Abstract

**Single photon emitters (SPEs) play an important role in a number of quantum information tasks such as quantum key distributions. In these protocols, telecom wavelength photons are desired due to their low transmission loss in optical fibers. In this paper, we present a study of bright single-photon emitters in cubic silicon carbide (3C-SiC) emitting in the telecom range. We find that these emitters are photostable and bright at room temperature with a count rate of ~ MHz. Together with the fact that SiC is a growth and fabrication-friendly material, our result may pave the way for its future application in quantum communication technology applications.**

## Introduction

Single-photon emitters (SPE) are used to generate "flying qubits" and they are critical in many quantum technology protocols such as quantum information processing[1-4], quantum simulation[5-7], quantum network and quantum communications[8-11]. Especially, in quantum key distribution (QKD), photons are used to send information in a secure way, protected by quantum mechanics[11]. Despite the fact that most QKD protocols use weak attenuated lasers to simulate SPE, a true SPE is still preferred due to its longer secure distance in theory and therefore its potentially better performance. In previous pioneer works, single photon quantum cryptography has been demonstrated with photons emitted from a single nitrogen-vacancy center in diamond[12]. The difficulty of QKD using true SPE lies on the fact that it is challenging to find a bright, room temperature (RT) SPE working in the telecom range, which is required to minimize the transmission loss in optical fibers.

Extensive efforts have been made to realize telecom-compatible SPE. InAs/InP QDs (quantum dots) present single photon emission at telecom wavelength[13] and they have been utilized to realize 120 km QKD[14]. However, QDs require operation at cryogenic temperatures, which makes their use experimentally more demanding. Most recently, room temperature SPEs in carbon nanotubes have equally been demonstrated to have emission in the telecom range[15, 16].

SiC is a wide band gap semiconductor widely used in LED industry. It is also a prominent material in the application of advanced high power, high temperature electronics. In recent years, defects in SiC have attracted increasing attention owing to their magneto-optical properties and the convenience for fabrication and scalability. Different types of SPEs have been discovered in SiC[17], such as carbon antisite–vacancy pair[18, 19], silicon vacancies[20-24], and divacancies[25-27]. However, those SPEs have emission either in the visible range[18-24, 28] or being weak in the near infrared range[25-27].

In this paper, we present a type of bright (~MHz) single emitters in 3C-SiC, which work at room temperature and emit in the telecom range. The sample we use is high-purity 3C-SiC epitaxy layer grown on a silicon substrate. First, we measured the photoluminescence (PL) spectrum of different SPEs and find that their fluorescence wavelengths lie in the telecom region. Then we investigated their optical properties: photo-stability and saturation behavior. Our results show that they have stable count rates of ~MHz at room temperature. Finally, we investigate their polarization properties for both excitation and emission, which demonstrate that these emitters can be treated as almost perfect single dipole. The polarization degree of both excitation and emission can reach up to around 97%. All these properties are highly desired in the QKD protocols with polarization coding scheme.

## Results

### Single photon emission

The use of an epitaxial layer ensures a low background level for SPEs detection. Moreover, it provides a natural way to obtain thin membranes for photonics and micromechanics applications rather than the thinning of bulk crystals[26]. We study the PL property of the emitters in home-made confocal microscopy systems. For the room temperature experiments, a 950 nm diode laser is used to excite the emitters through a high-NA (1.35) oil objective (Nikon). After passing through a 1000-nm dichroic mirror and a 1000-nm long pass filter, the florescence from the emitters is collected by a single-mode fiber. The emission is then guided to two channel superconducting single-photon detectors (SSPD, Scontel) for a Hanbury-Brown and Twiss (HBT) setup. In the cryogenic experiments, we use a closed cycle cryostation (Montana Instruments) combined with a confocal microscopy system with an infrared air objective with NA of 0.65.

Figure 1a shows a confocal scan of the single emitters in an area of $25 \times 25$ µm$^2$. Four bright SPEs are circled, which are verified by the continuous wave (CW) second-order autocorrelation function measurement using HBT interferometry. Then we continue to study the PL spectrum of the SPEs. As shown in Figure 1b, the center wavelengths of three representative RT PL spectra are 1085nm, 1188nm, and 1250nm. By measuring the PL spectrum of many SPEs, we find that the center wavelength of different emitters ranges from 1080nm to 1265 nm and the FWHM of the linewidth ranges from 128 nm to 194 nm. Figure 1c displays a low-temperature (4 K) PL spectrum of the two representative SPEs. There is no sizable reduction of the linewidth, which shows that the linewidth broadening is not caused by phonons. In one of the SPEs, several sharp lines appear in low temperature, which might come from emission coupled to phonons.

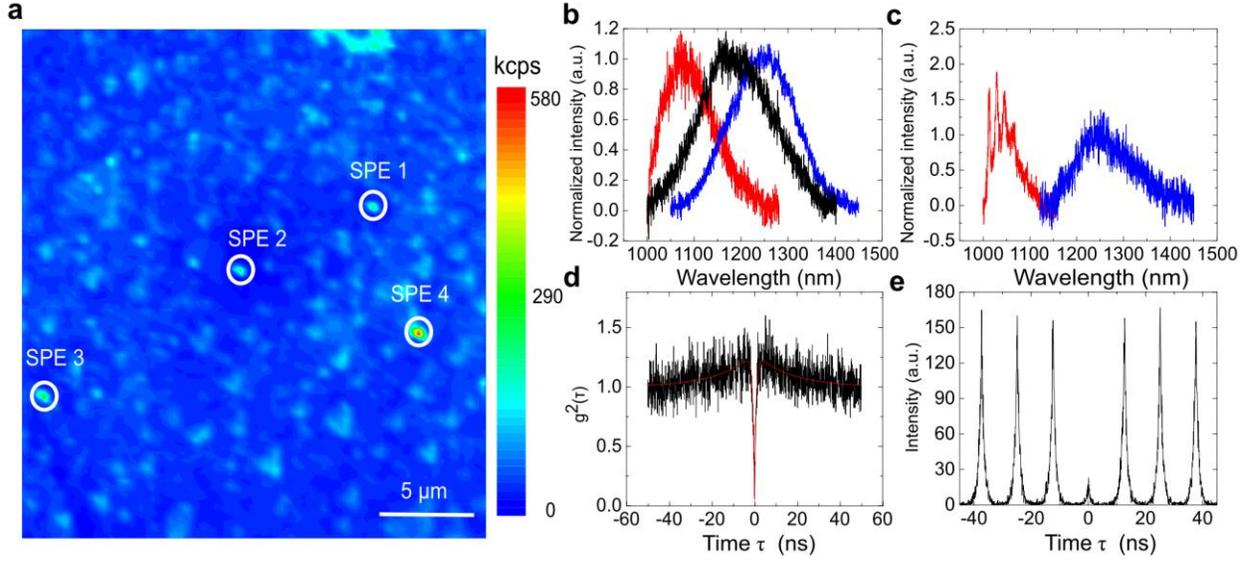

.

**Figure 1. 3C-SiC single photon emitters. (a)** 25 × 25 µm² confocal can map of the SPEs found in 3C SiC epitaxy layer under 2 mW laser excitation. The scale bar is 5 µm. SPEs are marked with white circles. **(b)** Three representative RT PL spectra of the SPEs. **(c)** Two representative PL spectra of the SPEs at cryogenic temperature (5 K). **(d)** Second-order autocorrelation function of SPE 4 with $g^2(0) = 0.05 \pm 0.03$ under 0.2 mW continuous wave (CW) laser excitation. The black line is the raw data and red solid line is the fitting with equation (1). **(e)** Second-order autocorrelation function of SPE 4 with $g^2(0) = 0.13 \pm 0.02$ under pulsed laser excitation (50 µW).

### Photon properties of the SPEs

Next we focus on one of the bright emitters and analyze its photon emission statistics with an HBT interferometer. Figure 1d shows the CW second-order autocorrelation function of SPE 4 (in Fig. 1a) under 0.2 mW excitation. The red line is the fitting result. In order to remove influence of the background, the raw autocorrelation function $g^2_{raw}(\tau)$ of the correlation function is corrected using the function $g^2(\tau) = [g^2_{raw}(\tau) - (1-\rho^2)]/\rho^2$ where ρ = s/(s+b). Here, s and b are the signal and background counts, respectively[21,26]. The background-corrected $g^2(\tau)$ is then fitted using the function

$$g^2(\tau) = 1 - (1+a)e^{-|\tau|/\tau_1} + ae^{-|\tau|/\tau_2} \qquad (1)$$

where a, τ₁, τ₂ are power dependent fitting parameters. From the fitting, the obtained value at zero time delay $g^2(0)=0.05\pm0.03$. Figure 1e shows pulsed second-order autocorrelation function measurement of SPE 4 using 950 nm picosecond laser (50 µW) with a repetition rate of 80 MHz. The obtained value at zero time delay $g^2(0)=0.13\pm0.02$. Which is well below 0.5 and therefore undoubtedly proves its nature of single photon emission. Other emitters show similar experimental results.

To further study the optical properties of the SPEs, we measured their lifetimes, power dependence and photo-stability. Figure 2a, 2b shows the spectrum and lifetime of the SPE 4. A lifetime of 0.81 ± 0.01 ns is obtained from the fitting with a single-exponential equation $I(t)\exp(-t/\tau)$ where $I(t)$ is the fluorescence intensity, and τ is the fluorescence lifetime. Next, we investigate the saturation behavior of the emitter SPE 4 under different power excitation. The total emission rate as a function of excitation power $P$ is presented in Figure 2c. The red line is the fitting of the date using the power dependence model $I(P)=I_s/(1+P_0/P)$, where $I_s$ is the maximal emission counts and $P_0$ is the saturation power. Inferred from the fitting, the maximum count rate $I_s$ is about 1.36 Mcps and the saturation power $P_0$ was about 3.0 mW. Moreover, fluorescence intensity trace of the emitter at the excitation powers of 1.5 mW and 8.2 mW with a sampling time bin size of 100 ms is displayed in Figure 2d. The emission is stable without photon blinking or photon bleaching even under a high power excitation (8.2 mW), which verified that it is a room temperature photostable SPE.

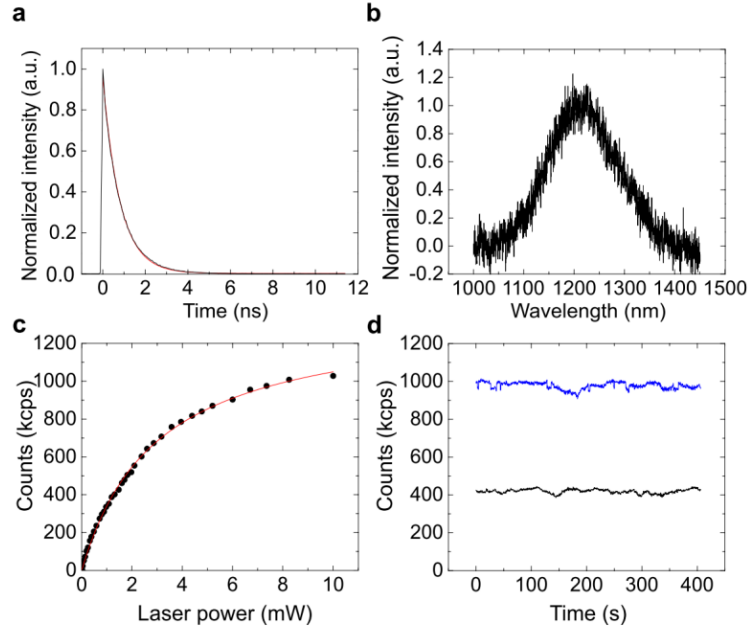

**Figure 2. Single photon emitter characterization.** (**a**) Lifetime measurement. The black curve is the raw data fitted with a single-exponential equation $I(t)\exp(-t/\tau)$. From the fitting, $\tau = 0.81 \pm 0.01 ns$ (**b**) RT PL spectrum of the SPE 4. (**c**) Saturation behavior of the SPE 4 at different laser powers. The black curve is the raw data fitted with $I(P) = I_s/(1+P_0/P)$. (**d**) Fluorescence intensity trace at laser powers 1.5 mW (black) and 8.2 mW (blue) with a sampling time of 100 ms and duration 400s at room temperature. No obvious blinking has been observed.

### Three-level model

In order to understand more about the energy levels of the emitter, we measure a set of power dependent auto-correlation functions, which can be modeled by a three-level model. Three representative g²(t) measurement are shown in figure 3b. The red line are the fittings of the data using the equation (1). In the measurement, an obvious power dependence of the photon-bunching effect is observed, which indicates that it has a metastable state[18-21]. Figure 3a shows a three-level model of the SPEs with ground state |1>, excited state |2>, and metastable state |3> and the pump-power-dependent transition rates, which is similar with other SPEs in SiC[18-21]. Figure 3 (c)-(e) shows the fitting parameters $a$, $\tau_1$ and $\tau_2$ of the power dependent g²(t) using the fitting equation (1). The parameters $a$ reveals the bunching, and $\tau_1$ reveals the transitions between the ground and excited state, $\tau_2$ reflects the behavior of the metastable state.

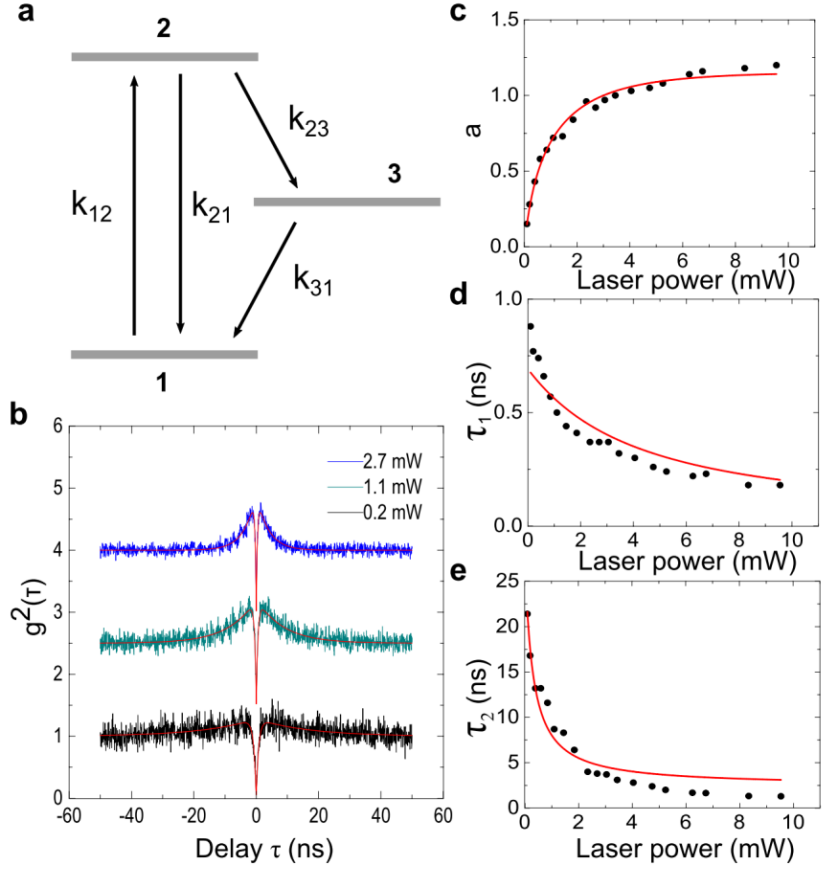

**Figure 3. Energy level analysis for SPEs.** (**a**) Three-level model of the SPEs with ground state |1>, excited state |2>, and metastable state |3>. The parameters $k_{12}$ is the excitation rate from ground state |1> to excited state |2>, and other $k_{ij}$ parameters represent the decay rate from state |i> to |j>, where i, j are number from 1 to 3. (**b**) Three representative CW second-order autocorrelation function measurements with different laser excitation powers. The red lines are the fitting using the equation (1). (**c**), (**d**), (**e**), The fitting parameters a, $\tau_1$ and $\tau_2$ of the antibunching curve as a function of excitation power. The red lines are the fitting of the data.

We fit the parameters using the functions based on three-level rate equation[21,29]

$$a = \frac{1-\tau_2 k_{31}}{k_{31}(\tau_2 - \tau_1)}, \tag{2}$$

$$\tau_{1,2} = \frac{2}{A \pm \sqrt{A^2 - 4B}}, \tag{3}$$

Where the $A = k_{12} + k_{21} + k_{23} + k_{31}$, $B = k_{12}(k_{23} + k_{31}) + k_{31}(k_{21} + k_{23})$. The fitting matches with the experimental data relatively well. Moreover, the calculated excited state lifetime, $(k_{21} + k_{23})^{-1}$, 0.7 ± 0.05 ns roughly matches the measured lifetime value 0.81 ± 0.02 ns.

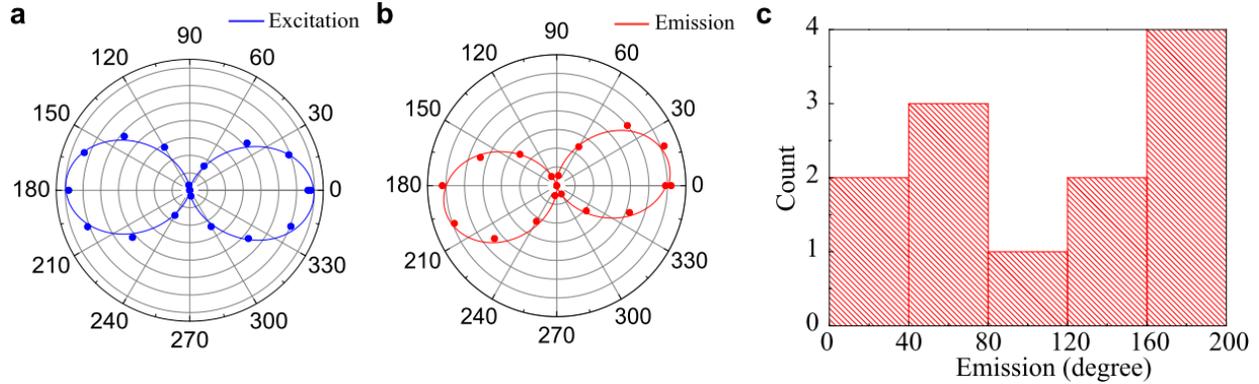

**Figure 4. Polarization of the SPEs. (a), (b),** The excitation (blue points) and emission (red points) polarization measurements from a SPE. Solid blue red lines are the fitting with $I = a + b\sin^2(\theta + \phi)$. (c) Summary of 12 emitters' emission axis orientation.

**Polarization measurement**

To know whether the SPE consists of a single dipole or multiple dipoles, we study the excitation and emission polarization by exciting the emitters using linearly polarized laser. The excitation polarization is measured by rotating a half-wave plate after the polarizer in the excitation path of the confocal setup while fixing the emission polarization measurements. The measurement of the emission polarization is similar to the method described above, but exchanges the role of excitation and collection arm. Figure 4a and 4b are the excitation (blue points) and emission (red points) polarization measurements from a SPE, respectively. The data are fitted by the function $I = a + b\sin^2(\theta + \phi)$, where $a, b, \theta$ and $\phi$ are the fitting parameters. The degree of the polarization ($\eta = (I_{max} - I_{min})/(I_{max} + I_{min})$) are about 97% and 98% for the excitation and emission polarization, respectively. The degree of both the excitation and emission polarization are larger than 90 % for all the studied emitters. These results show that these emitters have single polarizations, which is characteristic of a single linearly polarized dipole transition. Figure 4c displays 12 emitters' emission axis orientation. There is no obvious preference for the emission dipole directions. The high polarization degree (>95%) emission shows that they are well suitable for polarization encoding QKD such as in the Bennett-Brassard protocol.

Finally, we discuss the origin of these emitters. We attribute the SPEs to the same type defect due to the following indicators. The SPEs' PL spectrum at both room and low temperature are similar. Moreover, they show similar linear polarization behavior and the matching of a three-level system. In order to confirm that the SPEs are indeed in the 3C-SiC epitaxy layer, three pieces of sample with 1m epitaxy layer are etched using the reactive ion etching (RIE) to the depths of 200 nm, 300 nm, and 1 μm respectively. We find that there are single emitters after 200 nm and 300 nm depth etching. However, no emitters are found after 1 μm etching. This confirms that the SPEs are indeed in the 3C-SiC epitaxy film. Moreover, we also irradiate the samples with 1Mev electrons at a range of fluences (1E15cm^-2 to 5E16 cm^-2). Then the sample was annealed under 800 ◦C in a vacuum at $2 \times 10^{-4}$ Pa for half an hour. There is no observation of degradation regarding photostability, as well as no changes regarding the PL spectrum center wavelength position and the emitter density (see more in the Supplementary Note 1 and Supplementary Figure 2). In addition, we have tested another wafer (2.7 μm epitaxy layer grown on Si substrate from Air Wafer Inc). Similar SPEs have been found (See Supplementary Figure 1 and Supplementary table 1 for detailed information). These results indicate that the SPEs are intrinsic defects in the 3C-SiC epitaxial layer. The wavelength range of emission shows that the emitters we discovered represent a type of emitter which differs from other reported SPEs in 3C-SiC in the visible spectral range[19]. In the infrared range, one of the SPEs in 3C-SiC is well known and has been characterized: the divacancy defect (so called Ky5/L3 defect)[26-27]. However, its low temperature PL spectrum shows a narrow zero phonon lines which differs from our results. In addition, another major difference is that the saturated counts of Ky5 defect is about 25 kcps [19] which is significantly smaller than the one reported here. Based on the fact that our emitters have a variable center wavelength (~100nm), we rule out the possibility that strain in the 3C-SiC can induce such big PL shift. Like recently found SPEs in gallium nitride[30], our emitters PL variance may be related to stacking faults which usually present in hetero-epitaxial 3C-SiC layers. The photoluminescence is the result of exciton recombination process of electrons and holes. The hole is tightly localized at the point defect while the electron's position is loosely localized due to the stacking faults. Thus, the binding energy of the exciton and PL wavelength are determined by the position of the point defect. We have performed PLE measurement by changing the excitation wavelength with the same power and record count rate (Supplementary Note 2 and Supplementary Figure 3). We found that the most efficient excitation wavelength is around 975 nm (1.27 eV) for a SPE whose PL centered at around 1275 nm (0.97 eV). Clearly further investigations are needed in order to exactly identify the defects at the origin of these SPEs.

**Discussion**

In conclusion, we present the observation of bright RT telecom SPEs in the 3C-SiC epitaxial films grown on silicon substrates. These SPEs are RT photostable and the saturation counts are in the range of ~ Mcps without requiring the formation of photonics structures. It can be expected that by using some photonics structures, such as solid immersion lenses[20], nanopillars[23], microdisks[31, 32] and circular bullseye grating[33] the emission rate will be further increased. Moreover, the SPEs may also be used for electrically driven single-photon emitting devices[28]. Our finding of bright RT photostability telecom SPEs in the 3C-SiC can pave the way for its further applications in practical quantum key distribution with SPEs.

**Methods**

**Fitting procedure** In order to fit the power dependent parameters a, $\tau_1$ and $\tau_2$ of the $g^2(t)$ (Fig. 3 c-e), we model the power-dependent rate $k_{12}$ and $k_{31}$ as $k_{12} = KP$ and $k_{31} = (A_1/(1+B_1/P)+C)$ respectively where $A_1, B_1, C$ and $K$ are power-independent and $P$ indicates power. We then optimize the values of $A_1, B_1, C, K, k_{21}$ and $k_{23}$ such that the theoretical model described by equation (1) and (2) fits the power dependent data. The value of the fitting parameters found are: $A_1 = 0.15 \pm 0.03 GHz$, $B_1 = 2.9 \pm 1.1 mW$, $C = 0.035 \pm 0.003 GHz$, $K = 0.38 \pm 0.05 GHz/mW$, $k_{21} = 1.22 \pm 0.07 GHz$, and $k_{23} = 0.22 \pm 0.03 GHz$. A and B can be expressed by

$A = 1.475 + 0.38P + 0.15/(1+2.9/P)$ and $B = 0.0836P + (0.057P + 0.216)/(1+2.9/P)$, with the value A and B, $\tau_1$ and $\tau_2$ deduced by $\tau_{1,2} = 2/(A \pm \sqrt{A^2 - 4B})$.

## Data availability

The data that support the findings of this study are available from the corresponding authors on request.

**Acknowledgments** We thank the discussion with Adam Gali. We acknowledge the support from the Singapore National Research Foundation through a Singapore 2015 NRF fellowship grant (NRF-NRFF2015-03) and its Competitive Research Program (CRP Award No. NRF-CRP14-2014-02), Singapore Ministry of Education (MOE2016-T2-2-077 and MOE2011-T3-1-005), A*Star QTE programme and a start-up grant (M4081441) from Nanyang Technological University.

**Author Contributions**

These authors contributed equally: Junfeng Wang, Yu Zhou.

J. W., Y. Z., Z. W. built the optical set-up and performed the optical measurements. J. W., Y. Z., A. R., H.J. v. B. performed the data analysis. X. L and J. Y. performed electron irradiation of the samples. J. W. and W. G designed the experiments and wrote the manuscript with contributions from all co-authors. All authors contributed to the discussion of the results.

**Competing interests** The authors declare no competing interests.

**Corresponding author** Correspondence to lxj0218@hit.edu.cn，vonbarde@insp.jussieu.fr or wbgao@ntu.edu.sg.